\documentclass[12pt]{article}
\usepackage[a4paper, margin=1in]{geometry}

\usepackage[english]{babel}
\usepackage{amsmath, amsfonts, amssymb, dsfont, graphicx, tabularx, adjustbox, graphics, bbm, mathrsfs, mathtools, multirow, nameref, natbib, setspace, thmtools, bm, algorithm, algpseudocode}
\usepackage{enumitem}
\usepackage[font=footnotesize,justification=centering]{caption}
\usepackage{subcaption}

\usepackage[normalem]{ulem}

\usepackage{hyperref}

\usepackage{tabularx,booktabs}

\DeclareMathOperator*{\argmin}{argmin}

\usepackage{authblk}

\begin{document}
\def\spacingset#1{\renewcommand{\baselinestretch}%
{#1}\small\normalsize} \spacingset{1}

\title{Joint modeling for learning decision-making dynamics in behavioral experiments}

\author[1]{Yuan Bian}
\author[2]{Xingche Guo}
\author[1,3]{Yuanjia Wang}
\affil[1]{Department of Biostatistics, Columbia University, New York, USA}
\affil[2]{Department of Statistics, University of Connecticut, Connecticut, USA}
\affil[3]{Department of Psychiatry, Columbia University, New York, USA}
\date{}

\maketitle

\begin{abstract}
Major depressive disorder (MDD), a leading cause of disability and mortality, is associated with reward-processing abnormalities and concentration issues. Motivated by the probabilistic reward task from the Establishing Moderators and Biosignatures of Antidepressant Response in Clinical Care (EMBARC) study, we propose a novel framework that integrates the reinforcement learning (RL) model and drift-diffusion model (DDM) to jointly analyze reward-based decision-making with response times. To account for emerging evidence suggesting that decision-making may alternate between multiple interleaved strategies, we model latent state switching using a hidden Markov model (HMM). In the ``engaged'' state, decisions follow an RL-DDM, simultaneously capturing reward processing, decision dynamics, and temporal structure. In contrast, in the ``lapsed'' state, decision-making is modeled using a simplified DDM, where specific parameters are fixed to approximate random guessing with equal probability. The proposed method is implemented using a computationally efficient generalized expectation-maximization (EM) algorithm with forward-backward procedures. Through extensive numerical studies, we demonstrate that our proposed method outperforms competing approaches across various reward-generating distributions, under both strategy-switching and non-switching scenarios, as well as in the presence of input perturbations. When applied to the EMBARC study, our framework reveals that MDD patients exhibit lower overall engagement than healthy controls and experience longer decision times when they do engage. Additionally, we show that neuroimaging measures of brain activities are associated with decision-making characteristics in the ``engaged'' state but not in the ``lapsed'' state, providing evidence of brain-behavior association specific to the ``engaged'' state.
\end{abstract}

\noindent%

\maketitle
\doublespacing
\allowdisplaybreaks

\section{Introduction}
Major depressive disorder (MDD) is a mood disorder characterized by negative emotions, anhedonia (i.e., a diminished ability to feel pleasure), and psychomotor symptoms \citep{nelson1981symptoms}, significantly contributing to disability \citep{whiteford2013global} and mortality \citep{cuijpers2004increased}. Currently, MDD affects approximately 280 million people worldwide \citep{who2023} and is recognized as the leading global cause of disability \citep{friedrich2017depression}. Depressed individuals exhibit cognitive deficits \citep{rock2014cognitive} and concentration problems \citep{rice2019adolescent}, along with abnormalities in reward processing and altered learning abilities \citep{pizzagalli2005toward}. Recent studies \citep[e.g.,][]{lawlor2020dissecting, guo2024hmm} have demonstrated that individuals with MDD take longer to make decisions compared to healthy controls.

Reinforcement learning (RL) \citep{sutton2018reinforcement} is widely used to model cognitive processes and characterize decision-making behaviors, assuming that individuals learn through trial and error to optimize long-term rewards by developing an optimal policy. While specific RL methodologies may vary, their core principle remains consistent: individuals form and maintain internal representations of expected value, make decisions based on these expectations, and update these representations using the discrepancy between the expected and actual outcomes, known as the reward prediction error. \cite{huys2013mapping} introduced a prediction-error RL approach to characterize decision-making through key behavioral phenotypes, while \cite{guo2024semiparametric} extended this to a more general semiparametric RL framework. 

These error-driven RL models primarily focus on decision patterns and assume that the decision-making model follows a generalized linear model with a softmax link function. Although these models effectively capture trial-by-trial decision behavior, they fail to account for the complex dynamics underlying decision-making processes \citep{pedersen2017drift}. Decision-making behavior is known to involve a speed-accuracy trade-off \citep{wickelgren1977speed}, which can be assessed through response times. For example, abnormally short response times may indicate lapses in attention, leading to random decisions. Ignoring concurrent changes in response times when studying decision behaviors may therefore lead to biased or incomplete conclusions.

In contrast, drift-diffusion models \citep[DDMs;][]{ratcliff1978diffusion} conceptualize each decision as the continuous accumulation of noisy evidence until a decision boundary is reached.
A key advantage of DDM is its ability to simultaneously model decision outcomes and their temporal dynamics, revealing underlying cognitive processes such as information processing speed. DDMs are widely used to explain observed patterns of decision-making choices and response times in various tasks and model cognitive processes in psychiatric disorders. For instance, \cite{white2010using} applied DDMs to compare participants with different anxiety levels using a lexical decision task, and demonstrated that DDM could reveal deficits in decision-making that are undetectable through traditional response time analyses. \citet{ratcliff2015individual} analyzed both numerosity and lexical decision tasks with DDMs to investigate individual differences in decision-making. 
However, standard DDMs cannot account for how decision-making processes (e.g., processing speed or response caution) evolve with learning over trials \citep{miletic2020mutual}.

To summarize, RL methods effectively model how behavior adapts across trials but lack a mechanistic account of the decision-making process and do not capture response time distributions. Conversely, DDMs describe how decisions emerge from evidence accumulation and account for response time distributions, but they often overlook the aspect of learning over time \citep{miletic2020mutual}. To address these limitations, recent studies have integrated RL and DDM into a unified reinforcement learning diffusion decision model (RL-DDM) framework \citep[e.g.,][]{pedersen2017drift, fontanesi2019reinforcement}.
This framework not only improves parameter recovery and stabilizes model estimates \citep{shahar2019improving} but also outperforms standalone RL or DDM approaches by jointly modeling learning and response times, thus offering a more comprehensive characterization of decision-making dynamics \citep{miletic2020mutual}. 
For instance, \citet{pedersen2017drift} employed RL-DDMs in a probabilistic selection task to examine the effects of stimulant medication in adults with attention-deficit/hyperactivity disorder (ADHD), and \citet{fontanesi2019reinforcement} demonstrated that a modified RL-DDM captured simultaneous improvements in both decision speed and accuracy during learning. Similarly, \citet{fontanesi2019decomposing} applied RL-DDMs to a probabilistic instrumental learning task to disentangle the components of decision making under uncertainty.

Furthermore, classical RL, DDM, and RL-DDM methods typically assume that individuals employ a consistent decision-making strategy. Recent research, however, indicates that decision-making often switches between multiple strategies \citep[e.g.,][]{Worthy2013heterogeneity,Iigaya2018,calhoun2019unsupervised}. \cite{Ashwood2022} and \cite{guo2024hmm} demonstrated that decision-making can alternate between ``engaged'' and ``lapsed'' strategies, which conceptually align with the notions of ``exploitation'' and ``exploration'' in RL \citep{sutton2018reinforcement}. Under the ``engaged'' strategy, individuals make decisions following a softmax model, while under the ``lapsed'' strategy, decisions are made based on a fixed probability, largely ignoring external stimuli. \cite{li2024dynamic} considered a similar setting with ``engaged'' and ``random'' strategies, where the engaged strategy follows a softmax model and the random strategy involves uniform guessing. Such strategy switches can be effectively modeled using a hidden Markov model (HMM). However, the information contained in observed action or decision data alone is often insufficient to reliably estimate the parameters of such mixture models \citep{shahar2019improving}.
Moreover, \cite{Iigaya2018}, \cite{dillon2024using}, and \cite{guo2024hmm} showed that distinct decision-making strategies were associated with different response times, emphasizing the importance of accounting for strategy-switching and response times in decision-making models. 

In this work, we propose an RL-HMM-DDM framework to enhance the modeling of reward-based decision-making dynamics by incorporating multiple strategies and response times. In the ``engaged'' state, subjects make decisions according to an RL-DDM method, while in the ``lapsed'' state, decisions are modeled using a DDM method with certain parameters fixed to approximate random guessing with equal probabilities. This framework improves upon existing RL-DDM methods by allowing for latent strategy switching, making it more reflective of real-world decision-making. We use an HMM with covariate-dependent transition probabilities to model decision-making strategy switching, extending the frameworks of \citet{Ashwood2022} and \citet{li2024dynamic}, where the transition probabilities are assumed to be constants. By incorporating response times into the modeling process, our framework also improves upon the RL-HMM method in \citet{li2024dynamic} and \cite{guo2024hmm}. Under a two-arm action setup, the softmax model can be viewed as a simplified realization of the DDM, and therefore, RL-HMM can be interpreted as a special case of RL-HMM-DDM.

\subsection{Probabilistic Reward Task in the EMBARC Study}
The Establishing Moderators and Biosignatures of Antidepressant Response for Clinical Care (EMBARC) study \citep{trivedi2016establishing} is a randomized clinical trial of MDD that also enrolled healthy control (CTL) participants. As part of the study, participants completed the probabilistic reward task \citep[PRT;][]{pizzagalli2005toward}, a computer-based task designed to assess reward learning and behavioral adaptation to reinforcement. Each participant completed a baseline PRT session consisting of two blocks of $100$ trials, separated by a $30$-second break. On each trial, a cartoon face with either a short or long mouth was displayed, and participants were asked to identify the stimulus by pressing one of two buttons. The PRT promotes reward-based learning by disproportionately rewarding correct responses for one stimulus. Specifically, correct identification of the short-mouth face (the rich stimulus) was rewarded more frequently than that of the long-mouth face (the lean stimulus). Participants were instructed to maximize their total rewards, with the understanding that not all correct responses would be rewarded. To increase task difficulty, the difference in mouth length was deliberately subtle, creating a perceptual challenge that often biases participants toward favoring the more frequently rewarded stimulus. This design captures the natural tendency to adjust behavior based on reinforcement contingencies rather than perceptual precision.

We apply our proposed method to PRT data with the goals of characterizing decision-making behavior, identifying differences between MDD and CTL groups, and examining associations with brain measures and clinical outcomes. Our framework produces key behavioral variables, including group engagement rates, individual engagement scores, and individual response times during engagement and lapses, to be defined in Section \ref{subsec: pemer}. Group engagement rates capture systematic decision-making differences between MDD and CTL participants, while engagement scores and response-time measures provide individual-level behavioral markers potentially linked to brain function and clinical outcomes. These patterns may characterize cognitive processes underlying MDD and inform treatment strategies to enhance patient outcomes. Finally, our approach aims to infer and replicate observed behavior, aligning with principles of inverse reinforcement learning \citep{ng2000algorithms,abbeel2004apprenticeship}, behavioral cloning \citep{torabi2018behavioral}, and imitation learning \citep{ross2010efficient}, without assuming that behavior reflects an optimal reward-maximizing policy.

The remainder of the paper is organized as follows. Section \ref{sec: m} introduces general RL and DDM methods, outlines our proposed RL-HMM-DDM framework, and details the parameter estimation algorithm. Section \ref{sec: ss} presents simulation studies to evaluate the performance of the proposed method. In Section \ref{sec: aes}, we demonstrate the application of the proposed methods to the EMBARC study. We conclude with a discussion in Section \ref{sec: d} and defer additional results and technical details to the Supplementary Material.

\section{Methods}\label{sec: m}
\subsection{Objective and Data}
Consider data from $N$ individuals, each observed over $J$ trials. For each subject $i=1,\ldots,N$ and trial $j=1,\ldots,J$, the dataset consists of the tuple $(S_{i,j}, A_{i,j}, T_{i,j}, R_{i,j})$, where $S_{i,j}$, $A_{i,j}$, $T_{i,j}$, and $R_{i,j}$ represent the random variables corresponding to the state, action, response time, and reward, respectively. Throughout the paper, we use lowercase letters $s$, $a$, $t$, and $r$ to denote the realizations of $S_{i,j}$, $A_{i,j}$, $T_{i,j}$, and $R_{i,j}$, respectively.

Let $\mathcal{H}_{i,j}=\{S_{i,k},A_{i,k},T_{i,k},R_{i,k}\}_{k=1}^{j-1}$ represent the observed trial history up to trial $j$ for subject $i$. We assume that the state $S_{i,j}$ is confined to a bounded state space, denoted as $\mathcal{S}$, and is generated from an arbitrary non-trivial distribution of transitions $\Pr(S_{i,j} |\mathcal{H}_{i,j})$. In the special case of the contextual bandit, the states $S_{i,j}$ are assumed to be generated independently, i.e., $\Pr(S_{i,j}|\mathcal{H}_{i,j})=\Pr(S_{i,j})$. The action $A_{i,j}$ is selected from a binary action space $\mathcal{A}$, where $\mathcal{A}=\{0, 1\}$. The response time $T_{i,j}$ represents the time it takes for subject $i$ to make a decision in trial $j$, and is assumed to be finite. The reward $R_{i,j}$ follows a reward-generating distribution $\Pr_j(R_{i,j}|A_{i,j},S_{i,j},\mathcal{H}_{i,j})$, with a bounded reward space. The reward-generating mechanism may be either stochastic or deterministic. Our objective is to model human decision-making behavior driven by immediate rewards in behavioral experiments, where rewards are provided at the end of each trial. Accordingly, we do not consider any downstream effects of these rewards.

\subsection{Reward Processing with Reinforcement Learning}\label{subsec: rprl}
In general, reward processing refers to how individuals perceive, evaluate, and seek rewards, such as money or social approval. Biological evidence supports the use of simple reward prediction error, comparing actual and expected reward, to update value functions associated with different decisions \citep{schultz1997neural, waelti2001dopamine,o2003temporal}, laying the foundation for modeling reward processing utilizing reinforcement learning (RL). Let $Q_{i,j}(a,s)=E(R_{i,j}|A_{i,j}=a,S_{i,j}=s)$ represent the expected reward or value of taking action $a$ in state $s$ at trial $j$ for subject $i$. The reward prediction error for this trial is then given by $R_{i,j}-Q_{i,j}(a,s)$. Following the Rescorla-Wagner equation \citep{rescorla1972theory}, each subject's value evolves across trials and is updated based on a weighted reward prediction error as 
\begin{equation*}
Q_{i,j+1}(a,s)=Q_{i,j}(a,s)+\beta\{R_{i,j}-Q_{i,j}(a,s)\},
\end{equation*}
which can also be expressed as a weighted combination of the expected reward and the actual reward from the previous trial as $Q_{i,j+1}(a,s)=(1-\beta)Q_{i,j}(a,s)+\beta R_{i,j}$, where $\beta\in(0,1)$ is the learning rate, quantifying the speed at which the value of a state-action pair is updated. A larger $\beta$ indicates faster learning, where decisions are heavily influenced by rewards from recent trials. Conversely, a smaller $\beta$ results in slower updates, leading to decisions that integrate information from a longer history of past trials.

\subsection{Decision Making with Reinforcement Learning-Drift-Diffusion Model}\label{subsec: dmrlddm}
In a standard RL, the binary action $A_{i,j}$ is modeled as:
\begin{equation}
\label{eq: softmax}
    \Pr\bigl(A_{i,j}=1|S_{i,j}, \mathcal{H}_{i,j}\bigr)  =\frac{1}{1+\exp{ \bigl\{-\rho Z_{i,j}(S_{i,j})\bigr\} }},
\end{equation}
where $\rho$ represents reward sensitivity and $Z_{i,j}(S_{i,j})\triangleq Q_{i,j}(1,S_{i,j})-Q_{i,j}(0,S_{i,j})$ is the contrast between the expected rewards for the two actions in state $S_{i,j}$. This formulation emphasizes that the decision-making probability depends on the reward contrast $Z_{i,j}$, but it does not fully capture the dynamic complexities of decision-making processes.

\begin{figure}[h!]
     \centering
     \begin{subfigure}[h!]{0.49\textwidth}
         \centering
         \includegraphics[width=\textwidth]{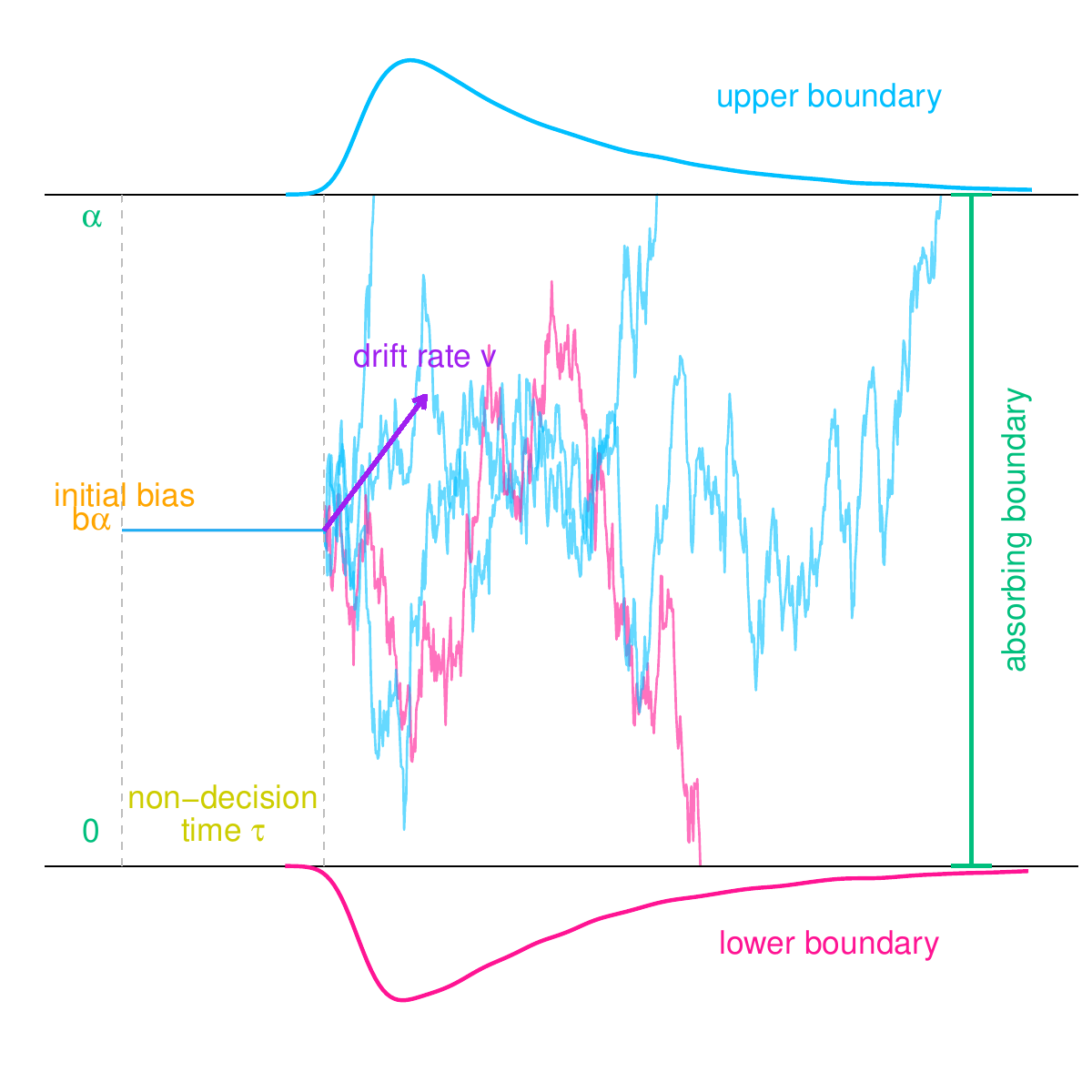}
         \caption{The drift-diffusion decision process.}
         \label{subfig: ddm}
     \end{subfigure}
     \hfill
     \begin{subfigure}[h!]{0.49\textwidth}
         \centering
         \includegraphics[width=\textwidth]{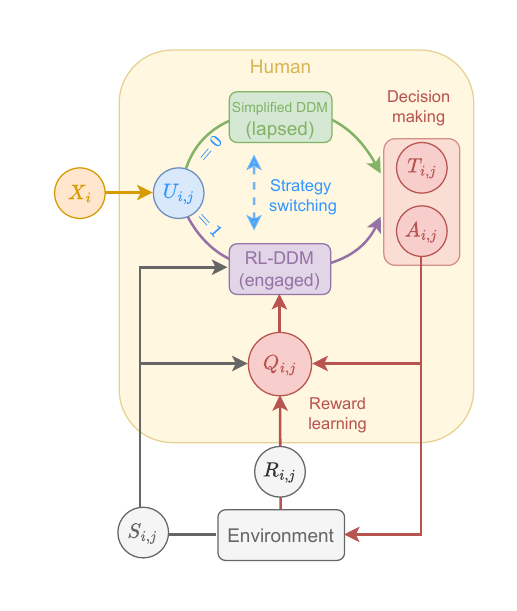}
         \vspace{-11mm}
         \caption{The RL-HMM-DDM framework.}
         \label{subfig: rlhmmddm}
     \end{subfigure}
     \vspace{-4mm}
     \caption{Graphical illustrations.}
\end{figure}

In contrast, the drift-diffusion model \citep[DDM;][]{ratcliff1978diffusion} conceptualizes decision-making as a continuous process of evidence accumulation. In this framework, the decision process begins at an initial state and proceeds until the noisily accumulated evidence reaches one of two absorbing boundaries corresponding to the available choices \citep{ratcliff2008diffusion}, as illustrated in Figure \ref{subfig: ddm}. Let $W_{i,j}(t)$ denote the evidence state at time $t$ for subject $i$ during trial $j$. The process is bounded by two absorbing boundaries: a lower boundary at $0$, yielding $A_{i,j}=0$, and an upper boundary at $\alpha$, yielding $A_{i,j}=1$, where $\alpha > 0$ governs the speed-accuracy trade-off. A larger value of $\alpha$ indicates higher accuracy, as the decision-making process takes longer to reach a decision. The initial evidence state $z \in (0,\alpha)$ reflects the initial bias, with the relative bias $b\triangleq z/\alpha$. A value of $b > 0.5$ indicates a bias toward the decision associated with $A_{i,j}=1$. The drift rate $v \in \mathbb{R}$ characterizes the direction and speed of evidence accumulation and is influenced by stimulus quality. When $v$ is close to zero, the stimulus is ambiguous; positive values of $v$ indicate evidence favoring $A_{i,j} = 1$, while negative values indicate evidence favoring $A_{i,j} = 0$.

The DDM describes $W_{i,j}(t)$ as a Wiener process with drift, such that its marginal distribution is $\text{Normal}(vt + b\alpha, \sigma^2 t)$. For simplicity, $\sigma$ is typically set to one  \citep[e.g.,][]{smith2000primer, navarro2009fast, blurton2012fast}. To convert to a scale where the variance parameter is fixed at any value $\sigma^2$, $\alpha$ and $v$ are rescaled by multiplying them by $\sigma$. The sub-densities \citep{horrocks2004modeling} of $T_{i,j}$ for $A_{i,j}=0$ and $A_{i,j}=1$ are denoted by $f_0(t;\alpha,b,v)\triangleq f(T=t,A=0;\alpha,b,v)$ and $f_1(t;\alpha,b,v)\triangleq f(T=t,A=1;\alpha,b,v)$, respectively. An example of these two densities is visualized in Figure \ref{subfig: ddm}. As suggested by their names, $f_0(t;\alpha,b,v)$ and $f_1(t;\alpha,b,v)$ are not probability density functions. However, the joint density function of $T_{i,j}$ and $A_{i,j}$, $f(t, a;\alpha,b,v)\triangleq\{f_0(t;\alpha,b,v)\}^{1-a}\{f_1(t;\alpha,b,v)\}^{a}$, forms a valid probability density function. 
Additional details on DDM are provided in Section S.1 of the Supplementary Material.

The response time $T_{i,j}$ consists of two components: the non-decision time $\tau$, which accounts for perceptual encoding and motor execution, and the decision time $T^{\text{D}}_{i,j} \triangleq \inf\{t : W_{i,j}(t) \leq 0 \text{ or } W_{i,j}(t) \geq \alpha\}$, which measures the time required for decision-making. The previous discussion represents a special case with a non-decision time of $\tau=0$ (i.e., the joint density is $f(t,a;\alpha,b,v,\tau=0)$). To incorporate a non-decision time $\tau > 0$, we substitute $t-\tau$ for $t$ in the function $f(t, a;\alpha,b,v)$; that is, \begin{equation}
\label{eq: wfpt}
f(t, a;\alpha,b,v,\tau)=f(t-\tau, a;\alpha,b,v,0),    
\end{equation} 
which yields the joint density function for the Wiener first-passage time (WFPT) distribution \citep{navarro2009fast}.
The probability of $A_{i,j}=1$ under such distribution is given by
\begin{align}
\label{eq: pa_ddm}
\Pr(A_{i,j}=1;\alpha,b,v,\tau)=\begin{cases}\dfrac{\exp(-2vb\alpha)-1}{\exp(-2v\alpha)-1} & \text{if } v\neq0,\\[1.5ex]
b & \text{if } v=0.
\end{cases}    
\end{align}

The DDM, however, cannot account for how decision-making processes (e.g., processing speed or response caution) evolve with learning over trials. To account for the negative association between $T_{i,j}$ and $Z_{i,j}$ observed in experiments \citep{frank2009prefrontal,krajbich2015rethinking}, an integrated framework known as the reinforcement learning drift-diffusion model \citep[RL-DDM;][]{pedersen2017drift} has been proposed. In this model, the drift rate varies across subjects and trials as $v_{i,j}=cZ_{i,j}$, with $c$ being a scaling parameter. This approach ensures that when the choice options are similar (and $Z_{i,j}$ is small), the drift rate $v_{i,j}$ is also small, thereby increasing the average time required to reach an absorbing boundary. Furthermore, when $b=0.5$ indicating that the RL-DDM is initially unbiased, \eqref{eq: pa_ddm} simplifies to:
\begin{align*}    
\Pr(A_{i,j}=1;\alpha,0.5,cZ_{i,j},\tau)= \frac{1}{1 + \exp(-\alpha cZ_{i,j})},
\end{align*}
which is equivalent to \eqref{eq: softmax} by setting $\rho = \alpha c$ \citep{tuerlinckx2005two, miletic2020mutual}. This equivalence reinterprets reward sensitivity $\rho$ as being proportional to decision caution through $\alpha$, demonstrating that RL-DDM generalizes standard RL with softmax model by integrating response times into modeling. However, when $b\neq0.5$, the equivalence between \eqref{eq: softmax} and \eqref{eq: pa_ddm} no longer holds, but \eqref{eq: pa_ddm} still approximates \eqref{eq: softmax} \citep{miletic2020mutual}.

\subsection{Strategies Switching with Hidden Markov Model}\label{subsec: sshmm}
The preceding development hinges on the assumption that individuals employ only one decision-making strategy, which is often violated in practice \citep[e.g.,][]{Worthy2013heterogeneity,Iigaya2018,calhoun2019unsupervised,Ashwood2022}. In this paper, we consider a framework where individuals alternate between two decision-making strategies, as illustrated in Figure \ref{subfig: rlhmmddm}. Let $X_i$ be a $p\times1$ vector of covariates for subject $i$, and let $U_{i,j}\in\{0,1\}$ denote a latent variable indicating the strategy used by subject $i$ in trial $j$. When $U_{i,j}=1$, individuals employ an ``engaged'' strategy, making decisions based on an RL-DDM model. Conversely, when $U_{i,j}=0$, they adopt a ``lapsed'' strategy, where decisions are modeled using a DDM model with $v$ and $b$ fixed at $0$ and $0.5$, respectively. When $v=0$ and $b=0.5$, \eqref{eq: pa_ddm} reduces to $\Pr(A_{i,j}=1)=0.5$, reflecting the probability of random guessing for each action. In Section S.2 of the Supplementary Material, we present a visualization of the DDM dynamics under the two decision-making strategies.

Within this behavioral decision-making framework, we assume that individuals' decision-making strategies are independent of how they learn from the actions they take and the rewards they receive. Even during the ``lapsed'' phase, individuals internally update their expectations of rewards, suggesting the presence of a computational mechanism for reward processing, known as implicit learning \citep{frensch2003implicit}. Engagement, however, depends on various external and internal factors such as mood, emotions, and potentially depression status, which may vary significantly between trials. 

By integrating the HMM and RL-DDM, we propose a novel RL-HMM-DDM framework to characterize perceptual decision-making behavior:
\begin{align}
f(T_{i,j},A_{i,j}|U_{i,j}, S_{i,j},\mathcal{H}_{i,j};\boldsymbol\vartheta)=&\ I(U_{i,j}=0)f(T_{i,j},A_{i,j};\alpha_0,0.5,0,\tau)\notag\\
&+I(U_{i,j}=1)f\{T_{i,j},A_{i,j};\alpha_1,b,c\cdot Z_{i,j}(S_{i,j}),\tau\},\label{eq: fta}\\
\Pr(U_{i,1}=k;\pi_k)=&\ \pi_k,\label{eq: ui1=k}\\
\Pr(U_{i,j+1}=1|U_{i,j}=k;\boldsymbol\varsigma)=&\ \frac{\exp{\left(\zeta_{k,0}+\zeta_{k,1}^{\top} X_{i}\right)}}{1+\exp{\left(\zeta_{k,0}+\zeta_{k,1}^{\top} X_{i}\right)}},\quad k=0,1,\label{eq: uij11}
\end{align}
where $f(T_{i,j},A_{i,j};\cdot,\cdot,\cdot,\cdot)$ is defined in \eqref{eq: wfpt}, and $\boldsymbol\vartheta = (\alpha_0, \alpha_1, b, c, \tau, \beta)^\top$ denotes the RL-DDM-specific parameters. The initial state probabilities satisfy $0<\pi_0,\pi_1<1$ with $\pi_0+\pi_1=1$. The transition model $\Pr(U_{i,j+1}=1|U_{i,j}=k;\boldsymbol\varsigma)$ is parametrized by $\boldsymbol\varsigma = \left(\zeta_{0,0}, \zeta_{0,1}^\top, \zeta_{1,0}, \zeta_{1,1}^\top\right)^\top$, where $\zeta_{k,0}$ and $\zeta_{k,1}$ represent the intercepts and slopes, respectively, for the logistic model in \eqref{eq: uij11}. The full set of model parameters is denoted by $\boldsymbol\theta = \left(\boldsymbol\vartheta^\top, \boldsymbol\varsigma^\top,\pi_1\right)^\top$.

\subsection{Parameter Estimation via EM Algorithm}\label{subsec: peema}
Rather than modeling the mechanisms by which states and rewards are generated, representing the participants' goals, we are interested in jointly modeling subjects' behaviors and decision times during reward-based decision-making. Consequently, the parameters in the state-generating model $\Pr(S_{i,j}|\mathcal{H}_{i,j})$ and reward-generating model $\Pr(R_{i,j}|S_{i,j},A_{i,j},\mathcal{H}_{i,j})$ are treated as nuisance parameters and considered independent of $\boldsymbol \theta$. This allows us to define the marginal likelihood function for $\boldsymbol\theta$, integrating out the latent decision strategy sequence $\boldsymbol U_i\triangleq U_{i,[1:J]}=(U_{i,1},\ldots,U_{i,J})^\top$, as:
\begin{equation*}
\prod^N_{i=1}\sum_{\boldsymbol U_i}\left\{\Pr(U_{i,1};\pi_1)\prod^{J-1}_{j=1}\Pr(U_{i,j+1}|U_{i,j};\boldsymbol\varsigma)\prod^J_{j=1}f(T_{i,j},A_{i,j}|U_{i,j}, S_{i,j}, \mathcal{H}_{i,j};\boldsymbol\vartheta)\right\}.    
\end{equation*}
We estimate $\boldsymbol \theta$ using a computationally efficient generalized expectation-maximization (EM) algorithm \citep{dempster1977maximum}, and let $\boldsymbol{\hat\theta}$ denote the resulting estimator. To implement the EM algorithm, we work with the complete-data log-likelihood:
\begin{equation*}
\sum^N_{i=1}\left\{\log \Pr(U_{i,1};\pi_1)+\sum^{J-1}_{j=1}\log\Pr(U_{i,j+1}|U_{i,j};\boldsymbol\varsigma)+\sum^J_{j=1}\log f(T_{i,j},A_{i,j}|U_{i,j}, S_{i,j}, \mathcal{H}_{i,j};\boldsymbol\vartheta)\right\}.
\end{equation*}
In the E-step, we compute the expected value of this complete-data log-likelihood under the current parameter estimates. To do this efficiently, we apply the forward-backward algorithm \citep{baum1970maximization} to compute the posterior distributions over the latent variables $U_{i,j}$ and their transitions. This dynamic programming approach leverages recursion and memorization, requiring only a single forward and backward pass through the trials for each subject. In the M-step, we update $\boldsymbol \theta$ by maximizing the expected complete-data log-likelihood with respect to the model parameters. The full procedure is outlined in Algorithm \ref{alg1}, with derivation details deferred to Section S.3 of the Supplementary Material.

\begin{algorithm}[h!]
\caption{RL-HMM-DDM algorithm}\label{alg1}
\begin{algorithmic}
\State Initialize $\boldsymbol\theta^{(0)}=\left\{\left(\boldsymbol\vartheta^{(0)}\right)^\top,\left(\boldsymbol\varsigma^{(0)}\right)^\top,\pi_1^{(0)}\right\}^\top$ and $\epsilon$, with $\boldsymbol\vartheta^{(0)}=\left(\alpha_0^{(0)}, \alpha_1^{(0)}, b^{(0)}, c^{(0)}, \tau^{(0)}, \beta^{(0)}\right)^\top$ and $\boldsymbol\varsigma^{(0)}=\left(\zeta_{0,0}^{(0)},\zeta_{0,1}^{(0)},\zeta_{1,0}^{(0)},\zeta_{1,1}^{(0)}\right)^\top$;
\For{iteration $m$ with $m=1,2,\ldots$}
    \State \textbf{E-step}: for subject $i$ with $i = 1, \ldots, N$ and $k=0,1$:
    \State \quad (i) update $\pi_0^{(m-1)}\gets1-\pi_1^{(m-1)}$, and update $\eta_{i,j,k}^{(m-1)}$ and $\varphi_{i,j,k,l}^{(m-1)}$ using $\boldsymbol\theta^{(m-1)}$;
    \State \quad (ii) \textbf{Initialize forward and backward variables}: $\rho_{i,1,k}^{(m)}\gets\pi_k^{(m-1)}\eta_{i,1,k}^{(m-1)}$ \quad and \quad $\varpi_{i,J,k}^{(m)}\gets1$;
    \State \quad (iii) \textbf{Forward recursion}: for $j=2,\ldots,J$,  $\rho_{i,j,k}^{(m)}\gets \eta_{i,j,k}^{(m-1)}\sum^1_{l=0}\rho_{i,j-1,l}^{(m)}\varphi_{i,j-1,l,k}^{(m)}$;  
    \State \quad (iv) \textbf{Backward recursion}: for $j=J-1,\ldots,1$, $\varpi_{i,j,k}^{(m)}\gets \sum^1_{l=0}\varpi_{i,j+1,l}^{(m)}\eta_{i,j+1,l}^{(m-1)}\varphi_{i,j,k,l}^{(m-1)}$; 
    \State \quad (v) \textbf{Posterior weights update}: for $j=1,\ldots,J$ and $l=0,1$,
    $$\gamma_{i,j,k}^{(m)}\gets\frac{\rho_{i,j,k}^{(m)}\varpi_{i,j,k}^{(m)}}{\sum^1_{l=0}\rho_{i,J,l}^{(m)}}\quad\text{and}\quad\xi_{i,j,k,l}^{(m)}\gets\frac{\rho_{i,j,k}^{(m)}\varpi_{i,j+1,l}^{(m)}\eta_{i,j+1,l}^{(m-1)}\varphi_{i,j,k,l}^{(m-1)}}{\sum^1_{l=0}\rho_{i,J,l}^{(m)}};$$
    \State \textbf{M-step}: 
    \State \quad (i) \textbf{Initial state probabilities update}:  $\pi_{1}^{(m)}\gets N^{-1}\sum^N_{i=1}\gamma_{i,1,1}^{(m)}$;
    \State \quad (ii) \textbf{HMM-specific parameters update}: for $k=0,1$, 
    \begin{align*}
    \left\{\zeta_{k,0}^{(m)},\left(\zeta_{k,1}^{(m)}\right)^\top\right\}\gets\argmin&\sum^N_{i=1}\sum^{J-1}_{j=1}\Big[-\xi_{i,j,k,1}^{(m)}\left(\zeta_{k,0}+\zeta_{k,1}^\top X_{i}\right)\\
    &+\left(\xi_{i,j,k,0}^{(m)}+\xi_{i,j,k,1}^{(m)}\right)\log\left\{1+\exp\left(\zeta_{k,0}+\zeta_{k,1}^{\top}X_{i}\right)\right\}\Big];
    \end{align*}
    \State \quad (iii) \textbf{RL-DDM-specific parameters update}: 
    \begin{align*}
    \boldsymbol\vartheta^{(m)}\gets\argmin &\sum^N_{i=1}\sum^{J}_{j=1} -\left(\gamma_{i,j,0}^{(m)}\log\{f(T_{i,j},A_{i,j};\alpha_0,1/2,0,\tau)\}\right.\\
    &+\left.\gamma_{i,j,1}^{(m)}\log[f\{T_{i,j},A_{i,j};\alpha_1,b,c\cdot Z_{i,j}(S_{i,j}),\tau\}]\right);
    \end{align*}
    \State \textbf{if} at iteration $m$,
    \begin{equation*}
    \left|\boldsymbol\theta^{(m)}-\boldsymbol\theta^{(m-1)}\right|\leq \epsilon
    \end{equation*}
    \State \textbf{then} stop iteration and define the final estimator as $\boldsymbol{\hat{\theta}}=\boldsymbol\theta^{(m-1)}$.
\EndFor
\end{algorithmic}
\end{algorithm}

For individual $i$, let $\boldsymbol A_i\triangleq A_{i,[1:J]}=(A_{i,1},\ldots,A_{i,J})^\top$ and $\boldsymbol T_i\triangleq T_{i,[1:J]}=(T_{i,1},\ldots,T_{i,J})^\top$ denote the sequences of actions and response times, respectively. At iteration $m$, define $\gamma_{i,j,k}^{(m)}=\Pr\left(U_{i,j}=k|\boldsymbol T_i,\boldsymbol A_i;\boldsymbol \theta^{(m)}\right)$ as the posterior marginal probability that subject $i$ used strategy $k$ at trial $j$, $\xi_{i,j,k,l}^{(m)}=\Pr\left(U_{i,j+1}=l,U_{i,j}=k|\boldsymbol T_i,\boldsymbol A_i;\boldsymbol \theta^{(m)}\right)$ as the posterior joint probability of a strategy transition from trial $j$ to $j+1$, $\eta_{i,j,k}^{(m)}=f\left(T_{i,j},A_{i,j}|U_{i,j}=k,S_{i,j},\mathcal{H}_{i,j};\boldsymbol \theta^{(m)}\right)$ as the joint likelihood of decision time and action under strategy $k$, and $\varphi_{i,j,k,l}^{(m)}=\Pr\left(U_{i,j+1}=l|\right.$ $\left.U_{i,j}=k;\boldsymbol \theta^{(m)}\right)$ as the transition probability from strategy $k$ to $l$. Define the forward variable $\rho_{i,j,k}^{(m)}=\Pr\left(U_{i,j}=k,T_{i,[1:j]},A_{i,[1:j]};\boldsymbol \theta^{(m)}\right)$ and the backward variable $\varpi_{i,j,k}^{(m)}= f\left(T_{i,[j+1:J]},\right.$ $\left.A_{i,[j+1:J]}|U_{i,j}=k,T_{i,[1:j]},A_{i,[1:j]};\boldsymbol \theta^{(m)}\right)$. These are computed recursively based on the Markov structure of the latent process. Specifically, the forward variables $\rho_{i,j,k}^{(m)}$ are initialized at $j=1$ using E-step (ii) of Algorithm \ref{alg1}, and updated for $j=2,\ldots,J$ according to E-step (iii); The backward variables $\varpi_{i,j,k}^{(m)}$ are initialized at $j=J$ using E-step (ii) and updated in reverse for $j=J-1,\ldots,1$ according to the E-step (iv); Using the normalized product of forward and backward variables, the posterior probabilities $\gamma_{i,j,k}^{(m)}$ and $\xi_{i,j,k,l}^{(m)}$ are computed in E-step (v). Given these expectations, the M-step updates the initial state probabilities, the HMM-specific parameters, and RL-DDM-specific parameters using M-steps (i), (ii), and (iii) of Algorithm \ref{alg1}, respectively.

\subsection{Post-Estimation Metrics for Engagement and Response Prediction}\label{subsec: pemer}
With $\boldsymbol{\hat\theta}$ estimated from Section \ref{subsec: peema}, we compute several quantities to evaluate individual engagement and predict decision-making behavior. Specifically, the individual engagement probability for subject $i$ at trial $j$, and the group-level engagement rate at trial $j$, are estimated as
\begin{equation*}
\hat\gamma_{i,j,1}=\Pr\left(U_{i,j}=1|\boldsymbol T_i,\boldsymbol A_i; \boldsymbol{\hat\theta}\right)
    \quad \mbox{and} \quad
    \hat\gamma_{j,1}= \frac{1}{N} \sum^N_{i=1}\hat\gamma_{i,j,1}.
\end{equation*}
$\hat\gamma_{i,j,1}$ serves as a predictive measure of of subject $i$'s engagement at trial $j$, while $\hat\gamma_{j,1}$ summarizes the average engagement probability across all individuals at trial $j$. Based on the individual engagement probability, we classify the latent strategy as \begin{equation}
\label{eq: upred}
\hat{U}_{i,j}=I(\hat{\gamma}_{i,j,1}\geq0.5).    
\end{equation}

To evaluate decision accuracy, define $\hat\eta_{i,j,k}\triangleq f\left(T_{i,j},A_{i,j}|U_{i,j}=k,S_{i,j},\mathcal{H}_{i,j};\boldsymbol{\hat\theta}\right)$, and $\hat\phi_{i,j,k}\triangleq f\left(T_{i,j},1-A_{i,j}|U_{i,j}=k,S_{i,j},\mathcal{H}_{i,j};\boldsymbol{\hat\theta}\right)$, where $\hat\eta_{i,j,k}$ denotes the likelihood of the observed action and response time, while $\hat\phi_{i,j,k}$ represents the likelihood of the observed response time paired with the opposite action. Then the posterior predictive probability of action $A_{i,j}$, given response time $T_{i,j}$, strategy $k$, current state $S_{i,j}$, and observed history $\mathcal{H}_{i,j}$, $\Pr\left(A_{i,j}|T_{i,j},U_{i,j}=k,S_{i,j},\mathcal{H}_{i,j};\boldsymbol{\hat{\theta}}\right)$, is estimated as
$\hat{\omega}_{i,j,k}=\hat\eta_{i,j,k}/\left(\hat\eta_{i,j,k}+\hat\phi_{i,j,k}\right)$. We define the predicted action as \begin{equation}
\label{eq: apred}
\hat{A}_{i,j}=I(\hat{\omega}_{i,j,k}\geq0.5).    
\end{equation} 
This measure allows us to assess how incorporating response times improves the identification and prediction of subjects’ actions.

To summarize engagement at the individual level, we define the individual engagement score as:
\begin{equation}
\label{eq: es}
\frac{1}{J}\sum^J_{j=1}
\log\left(\frac{\hat\gamma_{i,j,1}}{1-\hat\gamma_{i,j,1}}\right),
\end{equation}
which represents the average logit-transformed engagement probability across trials. We also compute the average response time during engagement and lapses,
\begin{equation}
\label{eq: rte}
\bigg\{\sum^J_{j=1}\mathbbm{1}\left(\hat{U}_{i,j}=k\right)\bigg\}^{-1} \sum^J_{j=1}\mathbbm{1}\left(\hat{U}_{i,j}=k\right)
T_{i,j},
\end{equation}
capturing the subject’s mean response time separately for trials classified as engaged when $k=1$ and lapsed when $k=0$, where $\mathbbm{1}(\cdot)$ denotes the indicator function.

\section{Simulation Studies}\label{sec: ss}
We assess the finite-sample performance of the proposed method through simulation studies. Two scenarios are considered: one that incorporates decision-making strategy switching and one that does not. Within each scenario, two settings are examined, each corresponding to a distinct reward-generating distribution. In setting 1, binary rewards are used, whereas setting 2 employs continuous rewards. Although the reward distributions differ across settings, the procedure for generating all other variables remains consistent. A total of $200$ simulations are conducted, using sample sizes $N\in\{100, 200\}$ and the number of trials $J \in \{100, 200\}$.

\subsection{Simulation Design and Data Generation}
We start by generating the state $S_{i,j}$ by flipping a fair coin. The covariates $X_i$ are set with $p=1$ and are independently drawn from Bern($0.6)$. Next, we simulate the latent decision-making strategy indicator $U_{i,1}$ from \eqref{eq: ui1=k} with $p_1=0.8$, and simulate $U_{i,j}$ for $j=2,\ldots,J$ from \eqref{eq: uij11} using parameters $\zeta_{0,0}=\zeta_{0,1}=-0.5$ and $\zeta_{1,0}=\zeta_{1,1}=1$. The action $A_{i,j}$ and response time $T_{i,j}$ are generated according to \eqref{eq: fta}, with parameters $\alpha_0=1$, $\alpha_1=1.5$, $b=0.6$, $c=2$, and $\tau=0.1$. The learning rate $\beta$ is set as $0.05$ and the starting reward expectation is specified as $Q_{i,1} = \bigl(\begin{smallmatrix}2 & 0 \\ 0 & 2\end{smallmatrix}\bigr)$.  For the reward $R_{i,j}$, we consider two different generating distributions depending on whether $A_{i,j}= S_{i,j}$. Specifically, if $A_{i,j}\neq S_{i,j}$, we set $R_{i,j}=0$. In setting 1, a Bernoulli distribution is used: when $A_{i,j}=S_{i,j}=1$, $R_{i,j}$ is generated from Bern($0.75$), representing a rich reward. Conversely, when $A_{i,j}=S_{i,j}=0$, $R_{i,j}$ is generated from Bern($0.3$), representing a lean reward. In setting 2, a Beta distribution is employed: when $A_{i,j}=S_{i,j}=1$, $R_{i,j}$ is generated from Beta$(3,1)$, corresponding to a rich reward. When $A_{i,j}=S_{i,j}=0$, $R_{i,j}$ is generated from Beta$(1,3)$ corresponding to a lean reward.

\subsection{Simulation Results}
Simulation results based on $200$ replicates for two settings are presented in Table \ref{tab: 1} for the scenario with decision-making strategy switching, and in Table \ref{tab: 2} for the scenario without strategy switching. We report the bias (Bias), empirical standard error (ESE), bootstrap standard error (BSE) based on $50$ bootstrap samples, and the coverage probability (CP) of the $95$\% confidence intervals, calculated under the normality assumption by $\mathrm{Est.} \pm 1.96 \times \mathrm{BSE}$. Our proposed RL-HMM-DDM method is compared against two alternatives: the RL-DDM method, which assumes a single ``engaged'' state across all trials, and the RL-HMM method, which uses a softmax model for decision-making and an HMM for strategy switching, while disregarding response times. Table \ref{tab: 1} indicates that when decision-making involves a mixture of two strategies, the RL-DDM method underestimates most parameters compared to RL-HMM-DDM. While the RL-HMM method also exhibits bias for the learning rate $\beta$ and HMM parameters, these results are not included here as this method does not estimate  $\alpha_0$, $\alpha_1$, $b$, $c$, and $\tau$. In contrast, when there is only one decision-making strategy, Table \ref{tab: 2} shows that both RL-HMM-DDM and RL-DDM perform well.

\begin{table}[h!]
    \centering
    \caption{Summary of the parameter estimates in 200 simulations with decision-making strategy switching}
    \label{tab: 1}
    \resizebox{\textwidth}{!}{\begin{tabular}{c c c r c c c r r c r r c c c r r c}
    \toprule
    & & & \multicolumn{7}{c}{Setting 1} & &  \multicolumn{7}{c}{Setting 2}\\
    \cmidrule{4-10}\cmidrule{12-18}
    & & & \multicolumn{4}{c}{RL-HMM-DDM} & & \multicolumn{2}{c}{RL-DDM} & & \multicolumn{4}{c}{RL-HMM-DDM} & & \multicolumn{2}{c}{RL-DDM}\\
    \cmidrule{4-7}\cmidrule{9-10}\cmidrule{12-15}\cmidrule{17-18}
    $N$ & $J$ & Parameters & \multicolumn{1}{c}{Bias} & ESE & BSE & CP & & \multicolumn{1}{c}{Bias} & ESE & & \multicolumn{1}{c}{Bias} & ESE & BSE & CP & &  \multicolumn{1}{c}{Bias} & ESE\\
    \midrule
    100 & 100 & $\beta$ & -0.0002 & 0.0024 & 0.0025 & 0.9600 & & -0.0111 & 0.0034 & & 0.0001 & 0.0022 & 0.0024 & 0.9650 & & -0.0102 & 0.0035\\
    & & $\alpha_1$ & -0.0011 & 0.0177 & 0.0182 & 0.9450 & & -0.3349 & 0.0079 & & -0.0025 & 0.0193 & 0.0176 & 0.9250 & & -0.3318 & 0.0084\\
    & & $b$ & -0.0000 & 0.0041 & 0.0040 & 0.9550 & & -0.0543 & 0.0030 & & -0.0003 & 0.0041 & 0.0041 & 0.9450 & & -0.0547 & 0.0032\\
    & & $c$ & -0.0027 & 0.0520 & 0.0514 & 0.9500 & & -1.0523 & 0.0407 & & -0.0012 & 0.0500 & 0.0504 & 0.9400 & & -1.0458 & 0.0409\\
    & & $\tau$ & 0.0001 & 0.0008 & 0.0009 & 0.9750 & & -0.0007 & 0.0008 & & 0.0000 & 0.0010 & 0.0009 & 0.9400 & & -0.0010 & 0.0009\\
    & & $\alpha_0$ & -0.0003 & 0.0090 & 0.0100 & 0.9700 & & & & & 0.0012 & 0.0090 & 0.0101 & 0.9700\\
    & & $\pi_1$ & -0.0002 & 0.0525 & 0.0554 & 0.9550 & & & & & -0.0010 & 0.0534 & 0.0552 & 0.9750\\
    & & $\zeta_{0,0}$ & 0.0127 & 0.1288 & 0.1214 & 0.9350 & & & & & 0.0152 & 0.1192 & 0.1239 & 0.9500\\
    & & $\zeta_{0,1}$ & -0.0112 & 0.1677 & 0.1584 & 0.9500 & & & & & -0.0154 & 0.1482 & 0.1624 & 0.9550\\
    & & $\zeta_{1,0}$ & -0.0007 & 0.1023 & 0.1016 & 0.9300 & & & & & -0.0056 & 0.1134 & 0.1047 & 0.9300\\
    & & $\zeta_{1,1}$ & 0.0051 & 0.1306 & 0.1302 & 0.9650 & & & & & 0.0135 & 0.1385 & 0.1330 & 0.9350\\
    100 & 200 & $\beta$ & 0.0000 & 0.0020 & 0.0020 & 0.9600 & & -0.0054 & 0.0035 & & 0.0002 & 0.0019 & 0.0020 & 0.9500 & & -0.0041 & 0.0036\\
    & & $\alpha_1$ & -0.0002 & 0.0113 & 0.0111 & 0.9300 & & -0.2798 & 0.0058 & & -0.0004 & 0.0111 & 0.0110 & 0.9500 & & -0.2771 & 0.0057\\
    & & $b$ & 0.0002 & 0.0035 & 0.0032 & 0.9300 & & -0.0588 & 0.0024 & & -0.0001 & 0.0031 & 0.0032 & 0.9600 & & -0.0595 & 0.0026\\
    & & $c$ & -0.0007 & 0.0399 & 0.0409 & 0.9550 & & -0.9468 & 0.0396 & &  0.0041 & 0.0418 & 0.0420 & 0.9550 & & -0.9309 & 0.0398\\
    & & $\tau$ & 0.0000 & 0.0007 & 0.0006 & 0.9250 & & -0.0053 & 0.0007 & & 0.0001 & 0.0007 & 0.0007 & 0.9500 & & -0.0055 & 0.0007\\
    & & $\alpha_0$ & -0.0000 & 0.0078 & 0.0077 & 0.9350 & & & & & -0.0010 & 0.0076 & 0.0078 & 0.9400\\
    & & $\pi_1$ & 0.0008 & 0.0549 & 0.0550 & 0.9550 & & & & & 0.0034 & 0.0545 & 0.0552 & 0.9650\\
    & & $\zeta_{0,0}$ & -0.0018 & 0.0852 & 0.0907 & 0.9600 & & & & & 0.0147 & 0.0957 & 0.0946 & 0.9350\\
    & & $\zeta_{0,1}$ & -0.0021 & 0.1186 & 0.1179 & 0.9600 & & & & & -0.0150 & 0.1256 & 0.1207 & 0.9400\\
    & & $\zeta_{1,0}$ & -0.0023 & 0.0799 & 0.0781 & 0.9500 & & & & & -0.0012 & 0.0808 & 0.0801 & 0.9400\\
    & & $\zeta_{1,1}$ & 0.0079 & 0.1028 & 0.0990 & 0.9500 & & & & & 0.0042 & 0.0956 & 0.1015 & 0.9600\\
    200 & 100 & $\beta$ & 0.0001 & 0.0017 & 0.0018 & 0.9600 & & -0.0108 & 0.0025 & & 0.0002 & 0.0017 & 0.0017 & 0.9700 & & -0.0102 & 0.0024\\
    & & $\alpha_1$ & 0.0013 & 0.0127 & 0.0128 & 0.9450 &  & -0.3341 & 0.0051 & & 0.0007 & 0.0139 & 0.0125 & 0.9200 & & -0.3314 & 0.0062\\
    & & $b$ & -0.0002 & 0.0030 & 0.0028 & 0.9400 &  & -0.0545 & 0.0023 & & -0.0001 & 0.0030 & 0.0029 & 0.9550 & & -0.0548 & 0.0021\\
    & & $c$ & 0.0039 & 0.0339 & 0.0361 & 0.9600 &  & -1.0509 & 0.0279 & & 0.0028 & 0.0376 & 0.0358 & 0.9150 & & -1.0461 & 0.0291\\
    & & $\tau$ & -0.0000 & 0.0006 & 0.0007 & 0.9750 & & -0.0008 & 0.0006 & & -0.0000 & 0.0007 & 0.0006 & 0.9750 & & -0.0011 & 0.0006\\
    & & $\alpha_0$ & -0.0003 & 0.0070 & 0.0070 & 0.9600 & & & & & -0.0002 & 0.0068 & 0.0071 & 0.9550\\
    & & $\pi_1$ & 0.0019 & 0.0369 & 0.0395 & 0.9650 & & & & & -0.0001 & 0.0391 & 0.0392 & 0.9500\\
    & & $\zeta_{0,0}$ & 0.0086 & 0.0782 & 0.0858 & 0.9750 & & & & & 0.0053 & 0.0824 & 0.0847 & 0.9550\\
    & & $\zeta_{0,1}$ & 0.0015 & 0.1060 & 0.1109 & 0.9550 & & & & & -0.0108 & 0.1121 & 0.1109 & 0.9600\\
    & & $\zeta_{1,0}$ & -0.0101 & 0.0692 & 0.0714 & 0.9700 & & & & & -0.0035 & 0.0691 & 0.0709 & 0.9550 \\
    & & $\zeta_{1,1}$ & 0.0044 & 0.0894 & 0.0919 & 0.9750 & & & & & 0.0062 & 0.0935 & 0.0915 & 0.9300\\
    200 & 200 & $\beta$ & 0.0002 & 0.0015 & 0.0014 & 0.9500 & & -0.0052 & 0.0023 & & 0.0002 & 0.0014 & 0.0014 & 0.9400 & & -0.0044 & 0.0025\\
    & & $\alpha_1$ & -0.0010 & 0.0080 & 0.0078 & 0.9400 & & -0.2796 & 0.0047 & & -0.0002 & 0.0078 & 0.0078 & 0.9450 & & -0.2771 & 0.0042\\
    & & $b$ & -0.0003 & 0.0023 & 0.0022 & 0.9350 & & -0.0589 & 0.0016 & & 0.0003 & 0.0024 & 0.0023 & 0.9400 & & -0.0594 & 0.0019\\
    & & $c$ & 0.0013 & 0.0301 & 0.0291 & 0.9450 & & -0.9427 & 0.0268 & & 0.0026 & 0.0309 & 0.0293 & 0.9250 & & -0.9352 & 0.0308\\
    & & $\tau$ & 0.0000 & 0.0005 & 0.0005 & 0.9300 & & -0.0053 & 0.0005 & & 0.0000 & 0.0005 & 0.0005 & 0.9150 & & -0.0056 & 0.0005\\
    & & $\alpha_0$ & -0.0002 & 0.0057 & 0.0055 & 0.9600 & & & & & -0.0004 & 0.0055 & 0.0055 & 0.9400\\
    & & $\pi_1$ & 0.0016 & 0.0333 & 0.0392 & 0.9800 & & & & & -0.0019 & 0.0373 & 0.0396 & 0.9700\\
    & & $\zeta_{0,0}$ & 0.0022 & 0.0615 & 0.0640 & 0.9700 & & & & & -0.0073 & 0.0635 & 0.0656 & 0.9550\\
    & & $\zeta_{0,1}$ & -0.0000 & 0.0810 & 0.0835 & 0.9600 & & & & & 0.0105 & 0.0819 & 0.0841 & 0.9450\\
    & & $\zeta_{1,0}$ & 0.0036 & 0.0568 & 0.0552 & 0.9450 & & & & & 0.0066 & 0.0541 & 0.0557 & 0.9550 \\
    & & $\zeta_{1,1}$ & -0.0013 & 0.0698 & 0.0699 & 0.9550 & & & & & -0.0047 & 0.0716 & 0.0710 & 0.9450\\
    \bottomrule
    \multicolumn{18}{l}{\footnotesize (Bias): estimate bias; (ESE): empirical standard error; (BSE): bootstrap standard error; (CP): coverage probability of the 95\% confidence intervals using BSE.}\\ 
    \multicolumn{18}{l}{\footnotesize (RL-HMM-DDM): our proposed method; (RL-DDM): RL-DDM method without decision-making strategy switching.}
    \end{tabular}}
\end{table}

\begin{table}[h!]
    \centering
    \caption{Summary of the parameter estimates in 200 simulations without decision-making strategy switching}
    \label{tab: 2}
    \resizebox{\textwidth}{!}{\begin{tabular}{c c c r c r r c r r c r r c}
    \toprule
    & & & \multicolumn{5}{c}{Setting 1} & &  \multicolumn{5}{c}{Setting 2}\\
    \cmidrule{4-8}\cmidrule{10-14}
    & & & \multicolumn{2}{c}{RL-HMM-DDM} & & \multicolumn{2}{c}{RL-DDM} & & \multicolumn{2}{c}{RL-HMM-DDM} & & \multicolumn{2}{c}{RL-DDM}\\
    \cmidrule{4-5}\cmidrule{7-8}\cmidrule{10-11}\cmidrule{13-14}
    $N$ & $J$ & Parameters & \multicolumn{1}{c}{Bias} & ESE & & \multicolumn{1}{c}{Bias} & ESE & & \multicolumn{1}{c}{Bias} & ESE & &  \multicolumn{1}{c}{Bias} & ESE\\
    \midrule
    100 & 100 & $\beta$ & 0.0001 & 0.0016 & & 0.0001 & 0.0016 & & 0.0002 & 0.0015 & & 0.0002 & 0.0015\\
    & & $\alpha_1$ & 0.0018 & 0.0100 & & -0.0005 & 0.0097 & & 0.0022 & 0.0103 & & -0.0004 & 0.0099\\
    & & $b$ & -0.0001 & 0.0027 & & 0.0003 & 0.0027 & & 0.0002 & 0.0026 & & -0.0001 & 0.0026\\
    & & $c$ & 0.0081 & 0.0297 & & 0.0009 & 0.0290 & & 0.0107 & 0.0321 & & 0.0032 & 0.0313\\
    & & $\tau$ & 0.0000 & 0.0010 & & 0.0001 & 0.0010 & & -0.0000 & 0.0009 & & 0.0000 & 0.0010\\
    100 & 200 & $\beta$ & 0.0002 & 0.0014 & & 0.0002 & 0.0014 & & 0.0001 & 0.0013 & & 0.0000 & 0.0013\\
    & & $\alpha_1$ & 0.0020 & 0.0069 & & 0.0001 & 0.0065 & & 0.0021 & 0.0065 & & 0.0003 & 0.0064\\
    & & $b$ & 0.0004 & 0.0020 & & 0.0001 & 0.0020 & & 0.0002 & 0.0020 & & -0.0000 & 0.0020\\
    & & $c$ & 0.0090 & 0.0256 & & 0.0014 & 0.0248 & & 0.0095 & 0.0246 & & 0.0020 & 0.0241\\
    & & $\tau$ & 0.0001 & 0.0008 & & 0.0001 & 0.0008 & & 0.0000 & 0.0008 & & 0.0000 & 0.0008\\
    200 & 100 & $\beta$ & -0.0001 & 0.0011 & & -0.0001 & 0.0011 & & 0.0001 & 0.0012 & & 0.0001 & 0.0012\\
    & & $\alpha_1$ & 0.0011 & 0.0076 & & -0.0007 & 0.0072 & & 0.0019 & 0.0074 & & 0.0004 & 0.0072\\
    & & $b$ & 0.0002 & 0.0019 & & 0.0001 & 0.0019 & & 0.0001 & 0.0020 & & -0.0001 & 0.0020\\
    & & $c$ & 0.0027 & 0.0227 & & -0.0029 & 0.0222 & & 0.0065 & 0.0241 & & 0.0017 & 0.0233\\
    & & $\tau$ & -0.0000 & 0.0007 & & 0.0000 & 0.0007 & & -0.0000 & 0.0007 & & 0.0000 & 0.0007\\
    200 & 200 & $\beta$ & 0.0001 & 0.0010 & & 0.0001 & 0.0010 & & 0.0000 & 0.0010 & & -0.0000 & 0.0010\\
    & & $\alpha_1$ & 0.0009 & 0.0045 & & -0.0007 & 0.0045 & & 0.0016 & 0.0046 & & 0.0001 & 0.0046\\
    & & $b$ & 0.0002 & 0.0014 & & -0.0000 & 0.0014 & & 0.0002 & 0.0015 & & -0.0000 & 0.0015\\
    & & $c$ & 0.0070 & 0.0190 & & 0.0009 & 0.0191 & & 0.0055 & 0.0188 & & -0.0005 & 0.0187\\
    & & $\tau$ & 0.0001 & 0.0006 & & 0.0001 & 0.0006 & & -0.0000 & 0.0006 & & -0.0000 & 0.0006\\
    \bottomrule
    \multicolumn{14}{l}{\footnotesize (Bias): estimate bias; (ESE): empirical standard error; (RL-HMM-DDM): our proposed method; (RL-DDM): RL-DDM}\\ 
    \multicolumn{14}{l}{\footnotesize method without decision-making strategy switching.}
    \end{tabular}}
\end{table}

\begin{figure}[h!]
    \centering
    \includegraphics[width=\linewidth]{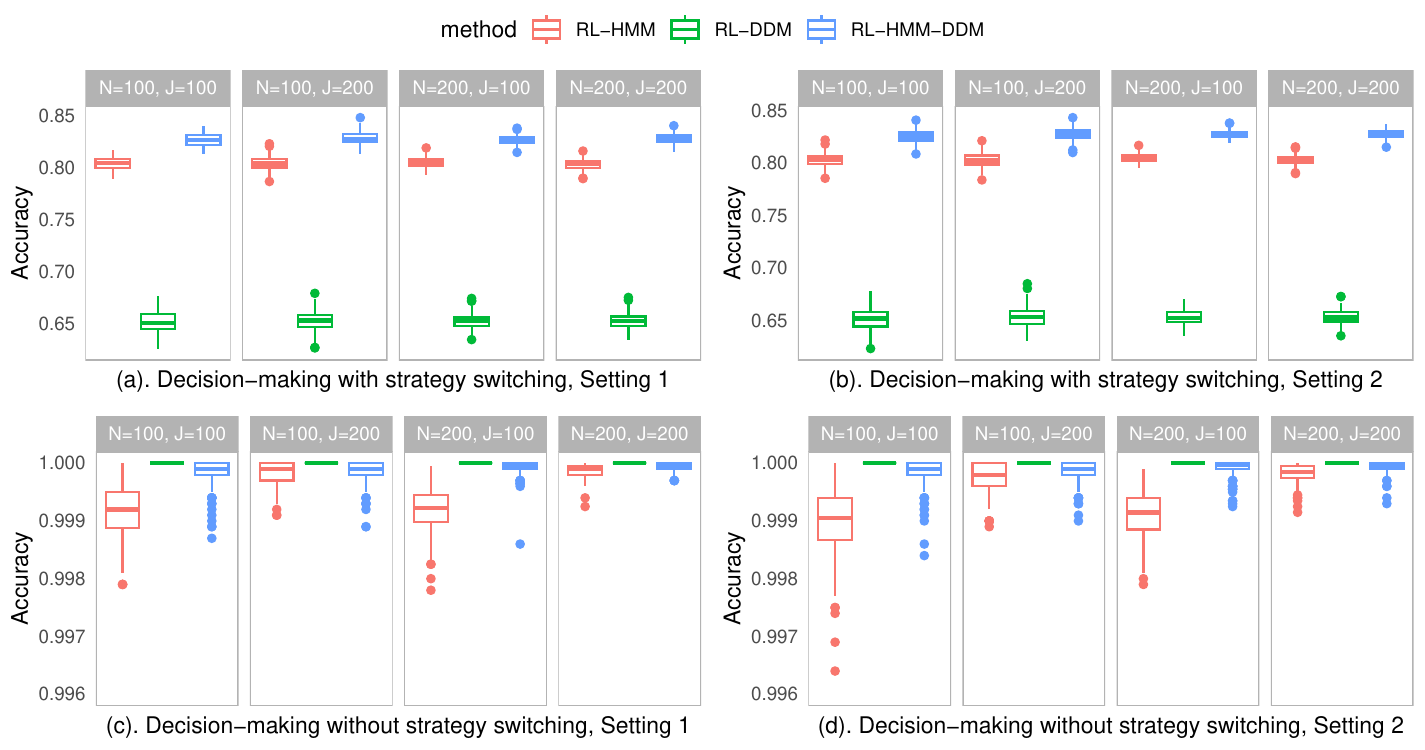}
    \vspace{-7mm}
    \caption{Estimation accuracy for $U_{i,j}$, summarized over the first 100 trials.}
    \label{fig: u}
\end{figure}

\begin{figure}[h!]
    \centering
    \includegraphics[width=\linewidth]{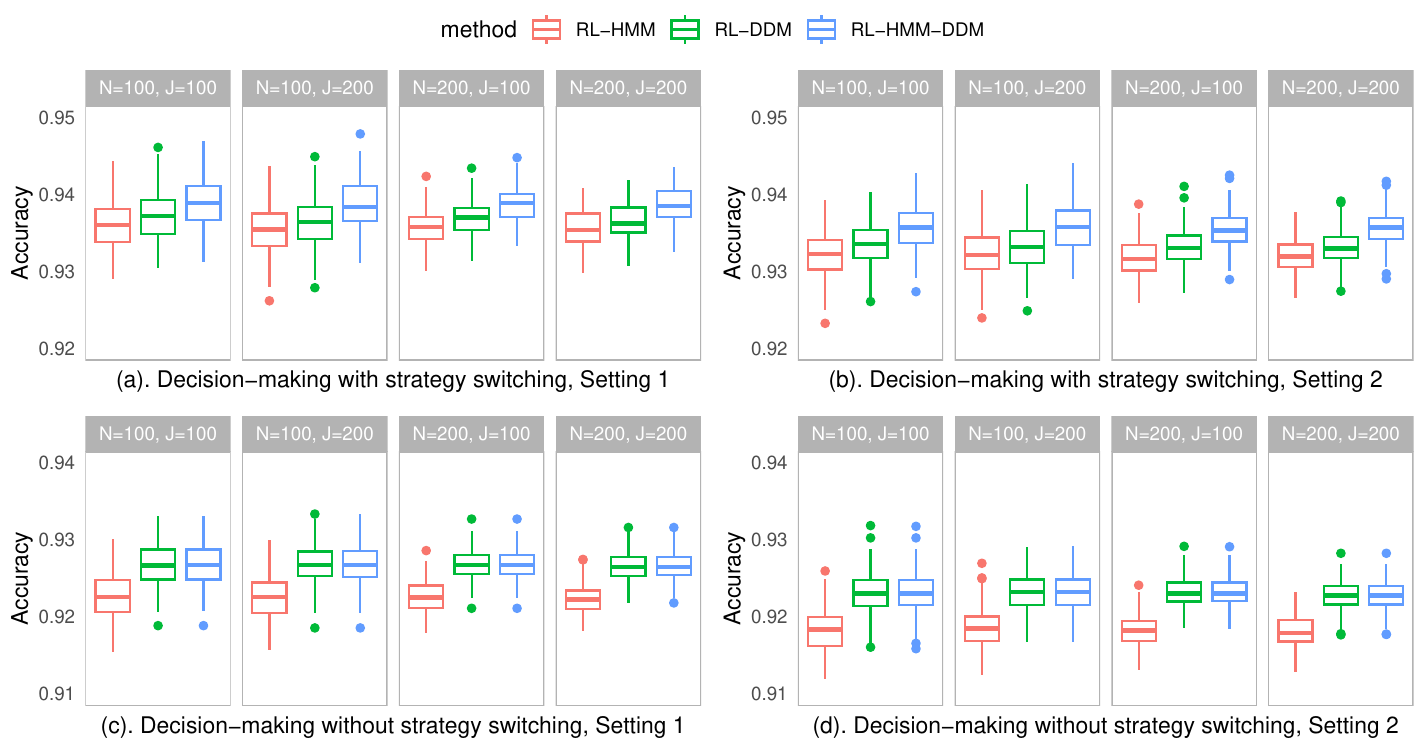}
    \vspace{-7mm}
    \caption{Estimation accuracy for $A_{i,j}$ when engaged, summarized over the first 100 trials.}
    \label{fig: a}
\end{figure}

We estimate the latent decision-making strategy and decision-making action using \eqref{eq: upred} and \eqref{eq: apred}, respectively. When $U_{i,j}=0$, decision-making is modeled as random guessing due to disengagement. To ensure a fair comparison across different methods and settings, we report results for $A_{i,j}$ only when $U_{i,j}=1$, representing trials in which the subject is engaged. Figures \ref{fig: u} and \ref{fig: a} summarize the results over the first $100$ trials for $U_i$ and $A_i$, respectively. 

In scenarios involving strategy switching, our proposed RL-HMM-DDM method achieves the highest estimation accuracies for both $U_{i,j}$ and engaged $A_{i,j}$ across all settings and combinations of sample size $N$ and number of trials $J$. By contrast, RL-HMM and RL-DDM achieve approximately $80$\% and $65$\% accuracy, respectively, in estimating $U_{i,j}$, over the first $100$ trials. For estimating engaged $A_{i,j}$ in the same window, RL-HMM performs the worst, while RL-DDM yields slightly better accuracy. In scenarios without strategy switching,  where RL-DDM represents the true model, RL-HMM-DDM and RL-HMM perform comparably to RL-DDM in estimating $U_{i,j}$, achieving accuracies above $99.80$\% and $99.60$\%, respectively. However, for estimating engaged $A_{i,j}$ over the first 100 trials, RL-HMM again shows the lowest accuracy, while RL-HMM-DDM and RL-DDM perform similarly. 

Furthermore, we compare the parameter estimates between RL-HMM-DDM and RL-HMM, and report F1 scores for $U_{i,j}$ in Section S.4 of the Supplementary Material, which support similar conclusions: RL-HMM-DDM outperforms the other methods in settings with strategy switching and performs comparably to RL-DDM in settings without strategy switching. In Section S.5 of the Supplementary Material, we evaluate the robustness of RL-HMM-DDM under input perturbations, such as noise in actions or response times, and find that RL-HMM-DDM is relatively robust to these contamination.

\section{Analysis of EMBARC Study}\label{sec: aes}
We analyzed the probabilistic reward task \citep[PRT;][]{pizzagalli2005toward} in the EMBARC study \citep{trivedi2016establishing}. Let $\mathcal{S}=\{0, 1\}$, where $0$ and $1$ represent lean and rich stimulus, respectively; and let $\mathcal{A}=\{0, 1\}$, where $0$ and $1$ correspond to the subjects selecting lean and rich stimulus, respectively. In a preliminary analysis, we observe that learning patterns may differ between the first and second blocks for subjects with MDD. To minimize potential bias, we focus on the first block with $J=100$. Next, we fit separate RL models for each subject, using the method described in Section \ref{subsec: rprl}, and exclude
participants with a learning rate less than $10^{-3}$. This results in $31$ subjects in the control (CTL) group and $153$ subjects in the MDD group, yielding a total of $N=184$ subjects. Let $X_i=0$ if the subject $i$ is in the CTL group, and $X_i=1$ if the subject $i$ is in the MDD group. 
Following \cite{ratcliff2008diffusion} and \cite{huys2013mapping}, we truncated extreme response times (RTs) to the range $[150,1500]$ms: any RT below $150$ms was set to $150$ms, and any RT above $1500$ms was set to $1500$ms.

\subsection{Comparisons Between MDD Patients and Controls}\label{subsec: cbmpc}
\begin{figure}[h!]
    \centering
    \includegraphics[width=\linewidth]{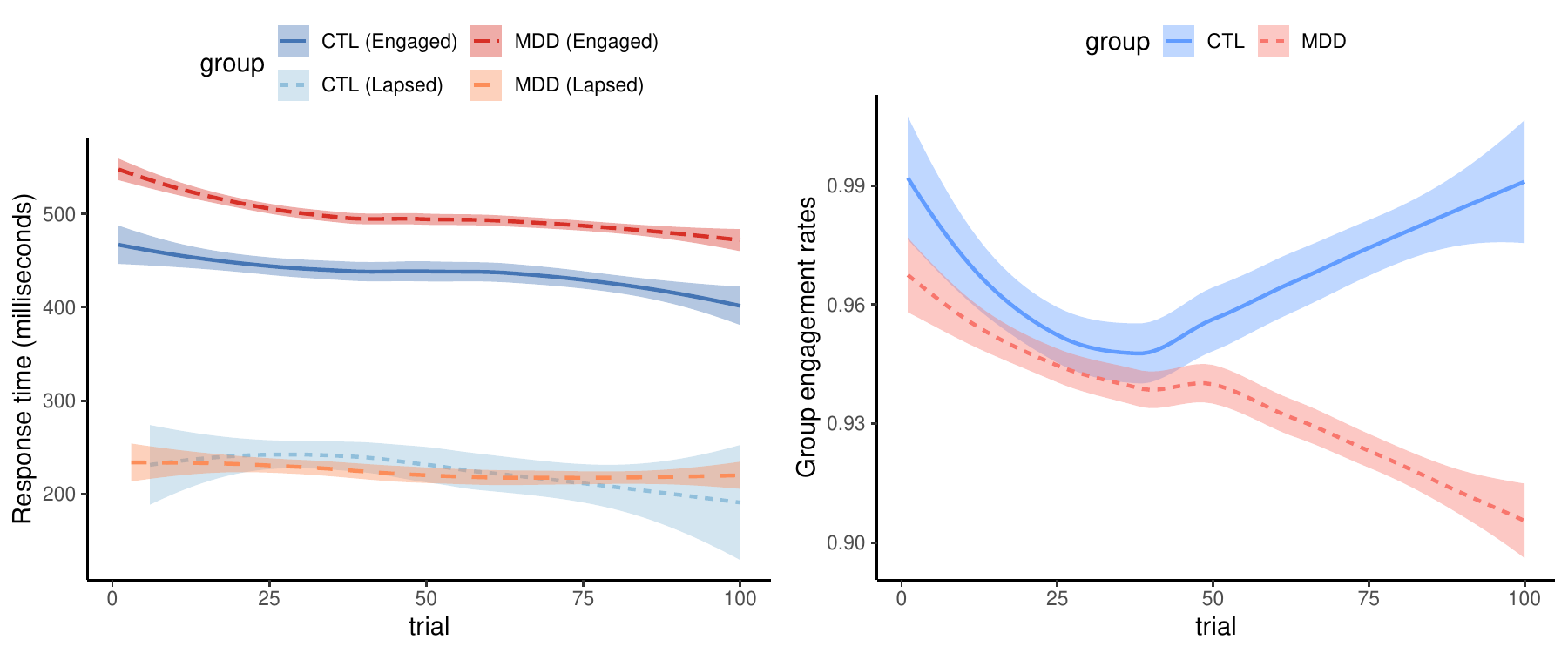}
    \vspace{-8mm}
    \caption{(Left) fitted response time trends for CTL and MDD in the engaged and lapse states; (Right) estimation of group engagement rates for CTL and MDD.}
    \label{fig: rtt}
\end{figure}

We analyze the data using the proposed method and present the results in Table \ref{tab: cc} and Figure \ref{fig: rtt}. To estimate the uncertainty, we generate 50 bootstrap samples to compute the bootstrap standard error (BSE) and construct a $95$\% bootstrap confidence interval (CI) for the parameters under the normality assumption. Given the skewness of $\pi_1$, we apply a logit transformation before using a normal approximation to construct the confidence interval. The correct decision rates are $78.49$\% and $51.71$\% for the estimated ``engaged'' and ``lapsed'' states, respectively, in the MDD group, and $81.41$\% and $45.92$\% in the CTL group. These findings indicate that the MDD group exhibits a lower correct decision rate in the ``engaged'' state compared to the CTL group. Additionally, in the ``lapsed'' state, the correct decision rates for both groups approach the level of random guessing. 

\begin{table}[h!]
    \centering
    \caption{Estimation of parameters in the PRT of the EMBARC study using the proposed method}
    \begin{tabular}{c r r c}
    \toprule
    Parameter & \multicolumn{1}{c}{EST} & BSE & 95\% CI\\
    \midrule
    $\beta$ & 0.035 & 0.003 & (0.028, 0.042)\\
    $\alpha_1$ & 1.343 & 0.039 & (1.267, 1.419)\\
    $b$ & 0.509 & 0.005 & (0.499, 0.519)\\
    $c$ & 2.984 & 0.007 & (2.970, 2.998)\\
    $\tau$ & 0.139 & 0.004 & (0.131, 0.147)\\
    $\alpha_0$ & 0.544 & 0.179 & (0.193, 0.895)\\
    $\pi_1$ & 0.999 & 0.013 & (0.985, 1.000)\\
    $\zeta_{0,0}$ & -1.326 & 3.318 & (-7.829,5.177)\\
    $\zeta_{0,1}$ & -0.770 & 2.558 & (-5.783,4.243) \\
    $\zeta_{1,0}$ & 4.851 & 0.517 & (3.837,5.865)\\
    $\zeta_{1,1}$ & -0.090 & 0.665 & (-1.393, 1.213)\\
    \bottomrule
    \multicolumn{4}{l}{\footnotesize (EST): estimate; (BSE): bootstrap standard error;}\\
    \multicolumn{4}{l}{\footnotesize (CI): confidence interval.}
    \end{tabular}
    \label{tab: cc}
\end{table}

Table \ref{tab: cc} reports a learning rate of $0.035$ and a relative bias of $b=0.509$, which is not statistically different from $0.5$. As discussed in Section \ref{subsec: dmrlddm}, when $b=0.5$, the RL-DDM reduces to a softmax model with reward sensitivity $\rho=\alpha_1c=4.01$. Those values are close to the ones reported in \cite{guo2024hmm}. The probability of being engaged in the first trial is $99.99$\% for all subjects. CTL subjects have a $99.22$\% probability of remaining in the ``engaged'' state if they were engaged in the previous trial, while this probability for MDD subjects is $99.15$\%. If subjects were in the ``lapsed'' state in the previous trial, they have a $20.98$\% probability of transitioning to the ``engaged'' state for CTL participants, and a $10.95$\% probability for those with MDD. However, these differences between MDD and CTL are not statistically significant.

Moreover, we use local polynomial regression to obtain the smoothed response times for MDD and CTL across trials for the two decision-making strategies, as well as the smoothed engagement rates for MDD and CTL. The estimated mean curves, along with the point-wise  $95$\% confidence bands of the mean curves, are displayed in Figure \ref{fig: rtt}. It illustrates that, on average, individuals with MDD take longer to make decisions when engaged compared to those in the CTL group. In contrast, when in the ``lapsed'' state, both groups exhibit similar, significantly shorter decision times compared to their ``engaged'' state. Furthermore, individuals in the CTL group are more likely to be in an ``engaged'' state, making deliberate decisions rather than relying on random guessing, compared to those with MDD. The engagement rate in the MDD group shows a non-increasing trend over trials, whereas in the CTL group, which consistently exhibits higher engagement than the MDD group, the engagement rate decreases during the first half of the trials and then increases in the second half.

We further separate response times by action and latent state-action pair in Section S.7 of the Supplementary Material. 
Both groups show decreasing response times for selecting the rich reward over trials, likely reflecting reinforcement learning and strategic adaptation, whereas response times for the lean reward remain stable. In the lapsed state, both groups show shorter, more uniform decision times, suggesting reduced cognitive engagement.

\subsection{Brain-Behavior Association in MDD Patients}\label{subsec: bba} Understanding the complex relationship between the brain and behavior is essential \citep{fonzo2019brain}. Investigating this connection may uncover the neural mechanisms underlying behavioral tasks, aiding in the identification of brain abnormalities associated with MDD, refining diagnostic frameworks, and facilitating the development of more effective therapeutic interventions.

\begin{figure}[h!]
    \centering
    \includegraphics[width=\linewidth]{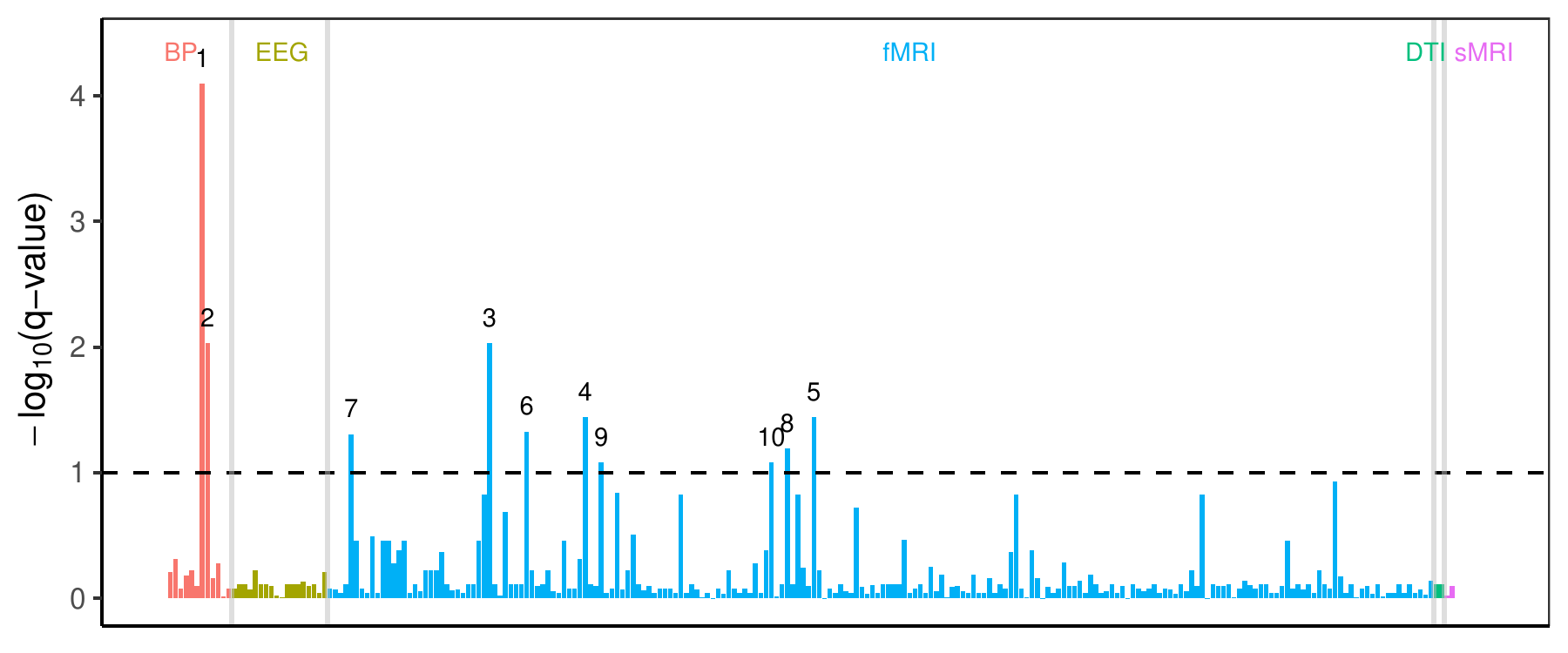}
    \caption{The $-\log10$ transformation of q-values of the regression coefficients for individual engagement scores regressed on various brain measures and clinical outcomes. The dashed line indicates the FDR at $10$\%.}
    \label{fig: er}
\end{figure}

\begin{figure}[h!]
    \centering
    \includegraphics[width=\linewidth]{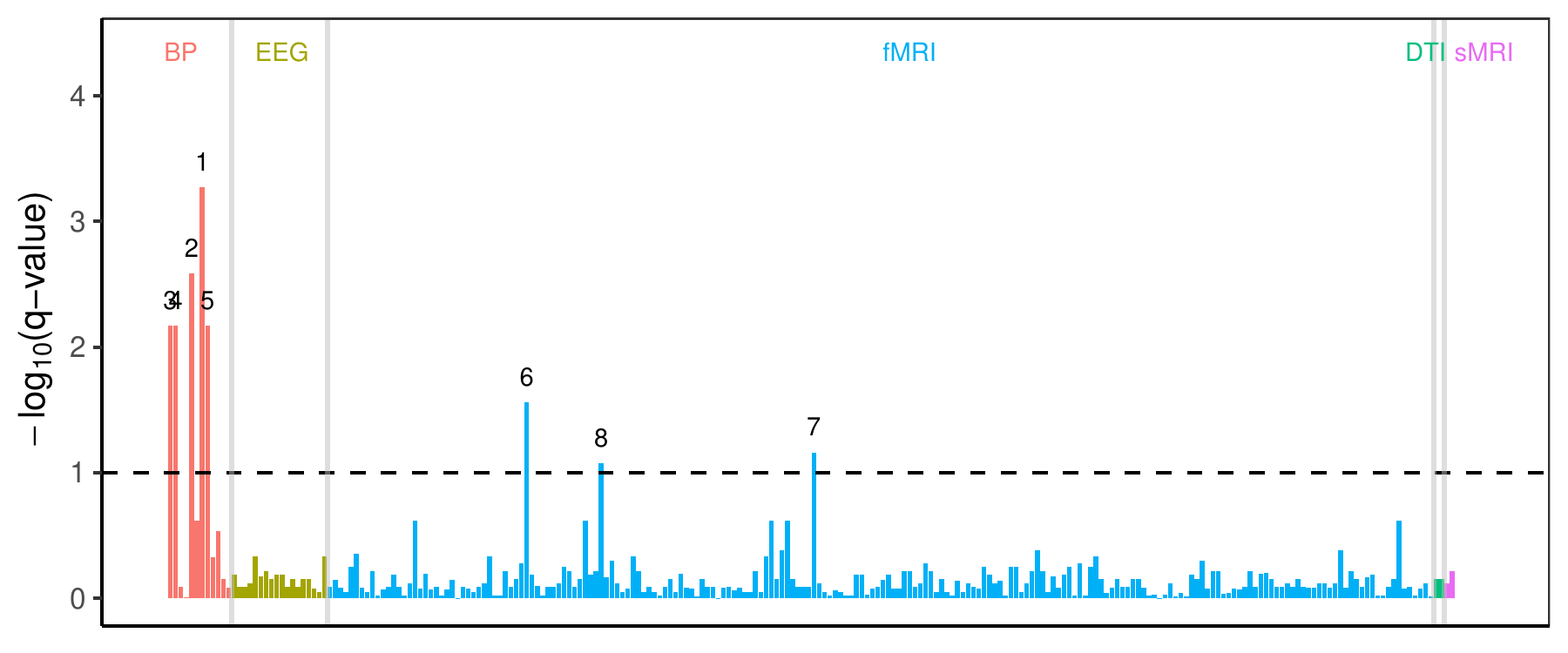}
    \caption{The $-\log10$ transformation of q-values of the regression coefficients for engaged response time regressed on various brain measures and clinical outcomes. The dashed line indicates the FDR at $10$\%.}
    \label{fig: ert}
\end{figure}

\begin{figure}[h!]
    \centering
    \includegraphics[width=\linewidth]{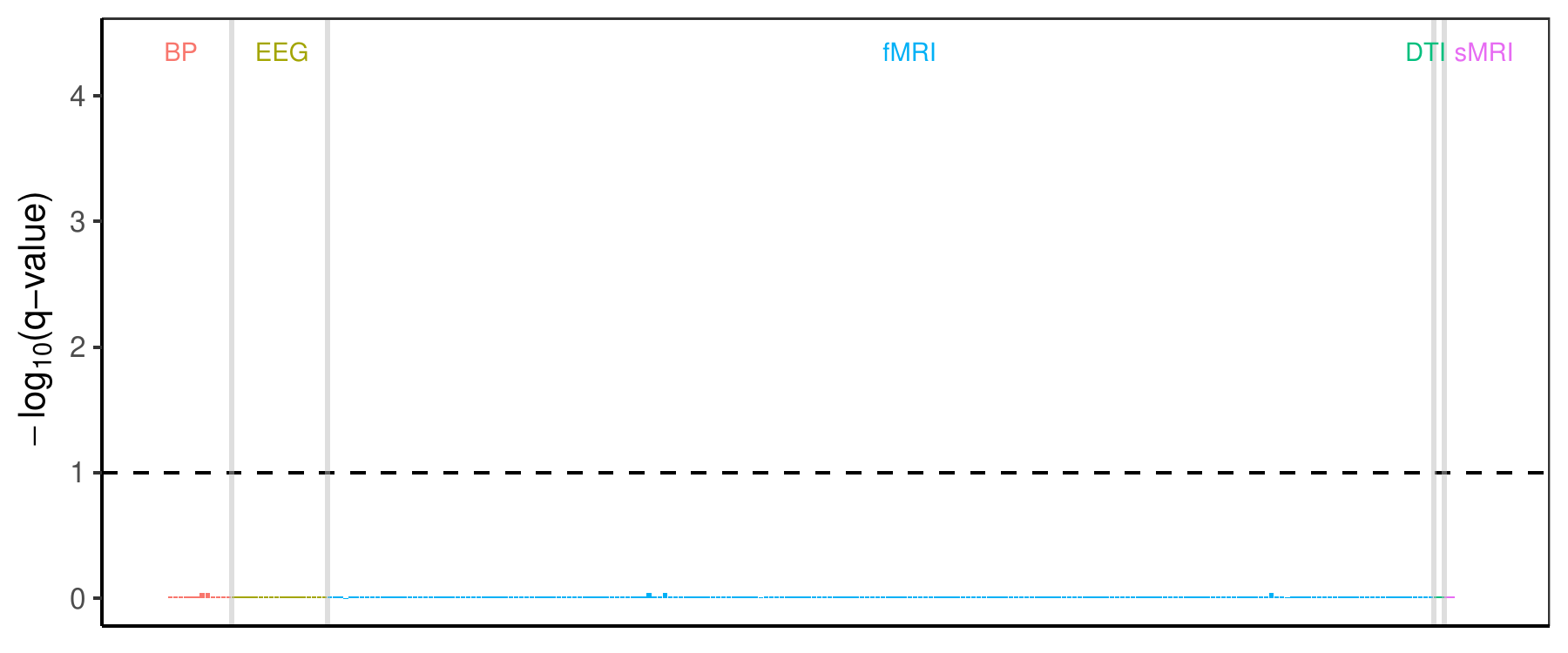}
    \caption{The $-\log10$ transformation of q-values of the regression coefficients for lapsed response time regressed on various brain measures and clinical outcomes. The dashed line indicates the FDR at $10$\%.}
    \label{fig: lrt}
\end{figure}

For individuals with MDD, we fit linear regression models to examine the relationships between three behavioral measures, i.e., individual engagement scores defined in \eqref{eq: es}, and response time during engagement and response time during lapses as defined in \eqref{eq: rte}, separately. We examined their associations with various biological variables, including behavioral phenotyping (BP), electroencephalogram (EEG), functional magnetic resonance imaging (fMRI), diffusion tensor imaging (DTI), and structural MRI (sMRI) measures. To control the false discovery rate (FDR), we compute p-values for the regression coefficients and apply q-value corrections \citep{benjamini1995controlling}. The results for engagement scores,  response times during engagement, and response times during lapses are summarized in Figures \ref{fig: er}, \ref{fig: ert}, and \ref{fig: lrt}. Details of the significant variables are provided in Tables \ref{tab: er} and  \ref{tab: ert}.

\begin{table}[h!]
    \centering
    \caption{Correlation estimates of significant variables with engagement rates}
    \resizebox{\textwidth}{!}{\begin{tabular}{l l l r c}
    \toprule
    \multicolumn{1}{c}{Modality} & \multicolumn{1}{c}{Task} & \multicolumn{1}{c}{Variable} & \multicolumn{1}{c}{EST} & 95\% CI\\
    \midrule
    BP & Flanker & accuracy & -0.457 & (-0.591, -0.298)\\
    BP & Flanker & reaction time & 0.353 & (0.181, 0.504)\\
    fMRI & Emotion Recognition & I minus C error mean & -0.339 & (-0.514, -0.137)\\
    fMRI & Resting First Block &  SCC / dorsal cingulate coupling & 0.388 & (0.201, 0.547)\\
    fMRI & Resting First Block & dACC / PCC coupling & 0.330 & (0.136, 0.499)\\
    fMRI & Resting First Block & left dlPFC / SCC coupling & 0.342 & (0.149, 0.509)\\
    fMRI & Resting First Block & left dlPFC / dACC coupling & 0.300 & (0.104, 0.474)\\
    fMRI & Resting First Block & right dlPFC / SCC coupling & 0.299 & (0.102, 0.473)\\
    fMRI & Resting First Block & right dlPFC / dACC coupling & 0.313 & (0.118, 0.485)\\
    fMRI & Resting First Block & right insula / dACC coupling & -0.347 & (-0.513, -0.155)\\
    \bottomrule
    \multicolumn{5}{l}{\footnotesize (EST): estimates; (CI): confidence interval.}
    \end{tabular}}
    \label{tab: er}
\end{table}

\begin{table}[h!]
    \centering
    \caption{Correlation estimates of significant variables with engaged response times}
    \resizebox{\textwidth}{!}{\begin{tabular}{l l l r c}
    \toprule
    \multicolumn{1}{c}{Modality} & \multicolumn{1}{c}{Task} & \multicolumn{1}{c}{Variable} & \multicolumn{1}{c}{EST} & 95\% CI\\
    \midrule
    BP & A Not B & median correct negative response time z-score & 0.344 & (0.175, 0.493)\\
    BP & A Not B & median correct total response time z-score & 0.345 & (0.176, 0.495)\\
    BP & Choice Reaction Time & median correct response time z-score & 0.372 & (0.209, 0.514)\\
    BP & Flanker & accuracy & -0.427 & (-0.566, -0.263)\\
    BP & Flanker & reaction time & 0.349 & (0.177, 0.501)\\
    fMRI & Resting First Block & dACC / PCC coupling & 0.344 & (0.152, 0.511)\\
    fMRI & Resting First Block & left dlPFC / dACC coupling & 0.305 & (0.109, 0.478)\\
    fMRI & Resting First Block & right insula / dACC coupling & -0.315 & (-0.486, -0.120)\\
    \bottomrule
    \multicolumn{5}{l}{\footnotesize (EST): estimates; (CI): confidence interval.}
    \end{tabular}}
    \label{tab: ert}
\end{table}

Significant correlations were observed between engagement scores and both BP and fMRI measures. Engagement scores are positively correlated with Flanker Task \citep{holmes2010serotonin}  reaction time ($r=0.353$), suggesting that higher engagement may be linked to slower but more deliberate responses in the Flanker Task, a pattern also observed in the PRT, as discussed in Section \ref{subsec: cbmpc}. These results align with previous studies highlighting a trade-off between cognitive control and response efficiency in other Flanker Tasks \citep{botvinick2001conflict}. Engagement scores are also negatively associated with Flanker Task accuracy ($r=-0.457$), suggesting that engagement in PRT may reflect a distinct process from cognitive control tasks. For example, reward pursuit in PRT and careful inhibition required for the Flanker Task may measure opposite cognitive or motivational processes. Positive associations between engagement scores and resting-state coupling are observed, particularly in regions such as the subgenual cingulate cortex (SCC) and dorsal cingulate ($r=0.388$), dorsal anterior cingulate cortex (dACC) and   posterior cingulate cortex (PCC) ($r=0.330$), and left / right dorsolateral prefrontal cortex (dlPFC) with SCC and dACC \citep{leech2014role}. Notably, right insula-dACC coupling shows a negative correlation with engagement ($r=-0.347$), possibly reflecting reduced salience-driven responses in highly engaged individuals, consistent with findings on salience network dynamics \citep{seeley2007dissociable, uddin2015salience}.

Engaged response times exhibit similar patterns. Flanker Task accuracy remains negatively correlated ($r=-0.427$), reinforcing the speed-accuracy trade-off. Engaged response times are positively associated with response times in the ``A Not B'' Task \citep{baddeley19683} and Choice Reaction Time Task \citep{thorne1985walter}. Additionally, engaged response times are positively associated with connectivity in the dACC-PCC ($r=0.344$) and left dlPFC-dACC ($r=0.305$) connections, suggesting that stronger prefrontal-cingulate coupling may support sustained cognitive control  \citep{dosenbach2008dual, cole2013multi}. Again, right insula-dACC coupling shows a negative relationship ($r=-0.315$), potentially reflecting alterations in salience network engagement, a phenomenon linked to cognitive flexibility and task-switching \citep{menon2010saliency, goulden2014salience}. These findings highlight the role of prefrontal and cingulate interactions in modulating cognitive effort and task engagement.

In contrast, response times in the ``lapsed'' state did not associate with any brain activity measures, as shown in Figure \ref{fig: lrt}. These analyses demonstrate the critical information gained by distinguishing the ``engaged'' versus ``lapsed'' state and accommodate different cognitive strategies used in each state, revealing how attention shapes both behavior and brain activity in decision making.

\section{Discussions}\label{sec: d}
In this paper, we propose a framework that integrates reinforcement learning (RL), hidden Markov model (HMM), and drift-diffusion model (DDM) to jointly model reward-based decision-making with response times. Our approach accounts for the switching of decision-making strategies between two states: an ``engaged'' state, where decisions follow an RL-DDM, and a ``lapsed'' state, where decisions are based on a simplified DDM, approximating random guessing. Extensive simulation studies demonstrate the robustness of our method across various reward-generating distributions, under both strategy-switching and non-switching scenarios, as well as in the presence of input perturbations. Applying our method to the EMBARC study yields novel findings of the decision-making process in individuals with MDD and CTLs. Specifically, RL-HMM-DDM shows that MDD has a lower overall engagement than CTLs and takes longer to make decisions when engaged. However, in the lapsed state, both groups demonstrate comparably shorter decision times than in the engaged state. Both groups show decreasing response times for selecting the rich reward over trials, likely reflecting reinforcement learning and strategic adaptation, whereas response times for the lean reward remain stable.
Furthermore, engagement scores and response times during engagement are associated with several brain measures and clinical outcomes, whereas response times during lapses are not. This finding underscores a brain-behavior association specific to the ``engaged" state and highlights the need for further investigation of mechanisms involving brain circuits such as PFC-ACC and insula-ACC.

Our method extends the DDM by incorporating an RL component into the drift rate $v$ of DDM in a linear manner. However, the framework could be further generalized to allow nonlinear integration and alternative RL-DDM formulations, e.g., the RL signal could be integrated into the absorbing boundary \citep{fontanesi2019reinforcement}. In our current model, random guessing is assumed as the decision-making strategy in the lapsed state. This assumption could be relaxed to accommodate alternative decision-making mechanisms, depending on the context or population under study. 

We also assume that the learning rates are identical across both the engaged and lapsed states. However, learning efficiency may be diminished in the lapsed state due to reduced cognitive engagement. To explore this, we conducted additional simulation studies presented in Section S.6 of the Supplementary Material. These simulations demonstrate that parameter estimates are accurate only when the true latent decision-making states are known. As misclassification of latent states increases, the estimates become increasingly unreliable. These findings underscore that the estimation of state-specific learning rates is highly sensitive to the accurate recovery of latent engagement states. Investigating this extension in greater depth remains an important direction for future research.

Another potential extension involves incorporating group-specific or subject-specific variations in RL-DDM parameters, such as the learning rate, absorbing boundary, initial bias, and non-decision time. Our methods are broadly applicable to analyzing behavioral tasks that include response time data. Furthermore, jointly modeling BP with other modalities, such as fMRI, presents a future direction for research. While this paper focuses on binary actions, the framework can be extended to accommodate multiple actions using race models \citep{marley1992horse}. These models conceptualize decision-making as a competition among parallel evidence-accumulation processes, with each option represented by a separate accumulator. The decision outcome is determined by the first accumulator to reach its threshold. Lastly, jointly model multiple behavioral tasks to distinguish between-subject associations and within-subject associations would be interesting.

\section*{Acknowledgments}
The authors thank the two reviewers, the Associate Editor, and the Editor for their valuable comments and suggestions, which have significantly improved the quality of this manuscript. This research is supported by U.S. National Institutes of Health grants MH123487, NS073671, and GM124104. Data used in the preparation of this manuscript were obtained from the National Institute of Mental Health (NIMH) Data Archive (NDA). NDA is a collaborative informatics system created by the National Institutes of Health to provide a national resource to support and accelerate research in mental health. Dataset identifier(s): DOI: 10.15154/fpa3-pd23. This manuscript reflects the views of the authors and may not reflect the opinions or views of the NIH or of the Submitters submitting original data to NDA.

\section*{Supplementary Material}
Supplementary Material Sections S.1-S.7 provide further technical details, additional simulation studies, and extended data-analysis results.

\bibliographystyle{apalike}
\bibliography{reference}

\end{document}